# Realism and Objectivism in Quantum Mechanics

## Vassilios Karakostas[∗]

**Abstract.** The present study attempts to provide a consistent and coherent account of what the world could be like, given the conceptual framework and results of contemporary quantum theory. It is suggested that standard quantum mechanics can, and indeed should, be understood as a realist theory within its domain of application. It is pointed out, however, that a viable realist interpretation of quantum theory requires the abandonment or radical revision of the classical conception of physical reality and its traditional philosophical presuppositions. It is argued, in this direction, that the conceptualization of the nature of reality, as arising out of our most basic physical theory, calls for a kind of contextual realism. Within the domain of quantum mechanics, knowledge of 'reality in itself', 'the real such as it truly is' independent of the way it is contextualized, is impossible in principle. In this connection, the meaning of objectivity in quantum mechanics is analyzed, whilst the important question concerning the nature of quantum objects is explored.

> When the layman says 'reality', he usually thinks that he is talking about something self-evident and well-known; whereas to me it appears to be the most important and exceedingly difficult task of our time to establish a new idea of reality.
>
> Pauli's letter to Fierz, 12.8.1948

## 1. Introduction

Throughout the development of physical science there has never been a theory which has changed so drastically the shape of science as quantum mechanics; nor has there been a scientific theory which has had such a profound impact on human thinking. Since its inception, quantum mechanics has played, and still does, a significant role in philosophical thought both as a source of metaphysical ideas and as an important example of a 'scientific revolution'. Thus, the advent of the quantum paradigm has gradually challenged the traditional philosophical substratum of science, the representational-visualizable description of microphysical entities and phenomena, the commonly perceived part-whole relationship that is built into classical physics, the concept of physical objects as carriers of completely determined properties, the unrestricted validity of deterministic laws, and even the nature of physical reality and its independence from the process of knowledge.

[∗] Department of Philosophy and History of Science, University of Athens, University Campus, Athens 157 71, Greece; e-mail: karakost@phs.uoa.gr
I wish to thank two anonymous referees of this journal for helpful comments and suggestions.

In such a perspective, it is often claimed, in investigations concerning the conceptual foundations of quantum mechanics, that the conception of realism is incompatible with quantum physics. This is the conclusion, with one qualification or another, that several investigators have drawn, especially on the basis of Bell's pioneering work. For instance, Michael Nielsen and Isaac Chuang in their book on the foundations of "Quantum computation and quantum information", after summarizing the consequences to be drawn from Bell's theorem, claim:

> What can we learn from Bell's inequality? … Most physicists take the point of view that it is the assumption of realism which needs to be dropped from our worldview in quantum mechanics, although others have argued that the assumption of locality should be dropped instead. Regardless, Bell's inequality together with substantial experimental evidence now points to the conclusion that either or both of locality and realism must be dropped from our view of the world if we are to develop a good intuitive understanding of quantum mechanics (Nielsen and Chuang 2010, p. 117).

While Arthur Fine in his much-discussed book "The shaky game", in its 2nd edition of 1996, expressly states:

> Realism is dead. … Its death was hastened by the debates over the interpretation of quantum theory, where Bohr's nonrealist philosophy was seen to win out over Einstein's passionate realism. Its death was certified, finally, as the last two generations of physical scientists turned their backs on realism and have managed, nevertheless, to do science successfully without it. To be sure, some recent philosophical literature has appeared to pump up the ghostly shell and to give it new life. I think these efforts will eventually be seen and understood as the first stage in the process of mourning, the stage of denial, … for realism is well and truly dead, and we have work to get on with, in identifying a suitable successor (Fine 1996, p. 112).

Setting within brackets the provoking mode in Fine's expression, I shall argue that it is not realism per se that is truly dead, it is rather the classical conception of scientific realism that requires a truly radical revision if a realist interpretation is to capture the content of quantum theory. Quantum features such as non-commutativity, non-separability and the generalized phenomenon of quantum entanglement have been forcing us to revise radically the intuitive classical ideas about physical reality. For a viable realist interpretation of quantum theory, the concept of realism must not be associated with ideas taken over from classical physics, such as atomism, localizability, separability, or similar philosophical preconceptions such as strict subject-object partition, mechanistic determinism and ontological reductionism.

At this point it is interesting to quote Hilary Putnam, who in his interview of 1994 with Burri emphatically states:

> The semantics underlying traditional realism are hopelessly metaphysical. In particular, numerous concepts of classical realism are untenable, for example the idea that one can reasonably talk about 'all entities' — as if the terms 'entity' or 'object' had a unique, fixed meaning — as well as the illusion that there is an answer to the question of which objects the world consists. The assumption that certain descriptions cover the world as it is in itself seems to be pointless to me (Burri 1994, p. 72).

Beyond Putnam's previous assertions, whose point of departure is a conviction of 'conceptual relativity' or a so-called 'internal realism' scheme (e.g., Putnam 1987), I would simply state that a number of views of traditional realist philosophy are incompatible with the results of modern science. Certain formulations of what traditional realism asserts are vague, so that it is difficult to evaluate their claims in the domain of science. Often such formulations are unnecessarily coupled with unfounded assumptions about the structure of the physical world. For example, it has been said that in a realistic interpretation the theoretical terms *genuinely* refer to objects existing in the world (e.g., Boyd 1983/1992, p. 195; Boyd 2002, sec. 5.3; see also Psillos 2000). Such a characterization, however, in its full generality, is inappropriate, since it implicitly suggests a specific assumption about the structure of the world, namely that the world consists or is built out of self-autonomous, intrinsically defined and independently existing objects. From the viewpoint of modern quantum theory, any *a priori* identification of 'physical objects' with 'physical reality' is inadmissible, since — whatever the precise meaning of 'physical objects' may be — we have to expect that such systems, according to the theory, are entangled by non-separable correlations of the EPR-type, so that they lack intrinsic individuality, intertemporal existence (see Sections 3.1, 3.2).

The non-classical nature of quantum systems surely presents a challenge to our understanding but this in no way implies resort to anti-realism. Anti-realist appreciations frequently arise from judging the concept of realism in quantum mechanics by classical or common-sensical standards. It is important, however, not to conflate the concept of physical realism with the specific sort of realism that may be extracted from classical physics (*classical physical realism* or *classical realism* for short; see Section 2.2). Such a conflation, in our view, still constitutes one of the serious 'epistemological obstacles' (in the sense of Bachelard 1938/1980) for preventing a genuine realist understanding of quantum mechanics. The process of overcoming obstacles of this kind presupposes, therefore, a principal differentiation of the key-conceptual assumptions underlying the classical and quantum mechanical worldview.

## 2. The Classical Conception of Nature

### 2.1. *Separability Principle and the Classical Ideal of Intelligibility*

Classical physics is essentially atomistic in character; it portrays a view of the world in terms of analyzable, separately existing but interacting self-contained parts. Classical physics is also reductionistic; a classical physical system can always be analyzed into its individual constituents, whose states and properties determine those of the whole they compose. Classical physics (and practically any experimental science) is further based on the Cartesian dualism of 'res cogitans' ('thinking substance') and 'res extensa' ('extended substance'), proclaiming a radical separation of an objective external world from the knowing subject that allows no possible intermediary.

In fact, the whole edifice of classical physics — be it point-like analytic, statistical, or field theoretic — is compatible with the following *separability principle* that can be expressed schematically as follows:

> *Separability Principle*: The states of any spatiotemporally separated subsystems $S_1$, $S_2$, ..., $S_N$ of a compound system $S$ are *individually well-defined* and the states of the compound system are *wholly* and *completely determined* by them and their spatiotemporal relations (Karakostas 2004, p. 284 and references therein).

In the case, for instance, of point-like analytic mechanics, the state of a compound system consisting of $N$ point particles is specified by considering all pairs $\{q_i(t), p_i(t)\}$, $i = 1, \ldots, 3N$, of the generalized position and momentum coordinates of the individual particles. Hence, at any temporal moment $t$, the individual pure state of the compound system consists of the $N$-tuple $\omega = (\omega_1, \omega_2, \ldots, \omega_N)$, where $\{\omega_i\} = \{q_i, p_i\}$ are the pure states of its constituent subsystems. It is then clear that in the individual, analytical interpretation of classical mechanics maximal knowledge of the constituent parts of a compound system provides maximal knowledge of the whole system (see, for example, Scheibe 1973, pp. 53-54). Accordingly, every property the compound system has at time $t$, if encoded in $\omega$, is determined by $\{\omega_i\}$. For instance, any classical physical quantities (such as mass, momentum, angular momentum, kinetic energy, center of mass motion, gravitational potential energy, etc.) pertaining to the overall system are determined in terms of the corresponding quantities of its parts. They either constitute direct sums or ordinary functional relations (whose values are well-specified at each space-time point) of the relevant quantities of the subsystems. Thus, they are wholly determined by the subsystem states. Furthermore, given the state $\omega_t(q, p)$ of a classical system in phase space at time $t$, the dynamical law which connects $\omega_t$ with the state $\omega_{t'}(q, p)$ of the system at any other time $t'$ is given by the Hamiltonian $H(q, p)$ and the canonical equations. This means that a classical system $S_t$, uniquely defined at time $t$, can be re-identified at any other time $t' \neq t$ by the phase point $(q_t, p_t)$ values on its dynamical trajectory. Hence, classical physics determines objects-systems as individuals with *intertemporal identity*. They can be identified through conservation of their essential quantities, re-identified in time, and distinguished from their like. The foregoing concise analysis delimits actually the fact, upon which the whole classical physics is founded, that any compound physical system of a classical universe can be conceived of as consisting of *separable*, *individual* parts interacting by means of forces, which are encoded in the Hamiltonian function of the overall system, and that, if the full Hamiltonian is known, maximal knowledge of the values of the physical quantities pertaining to each one of these parts yields an exhaustive knowledge of the whole compound system, in perfect conformity with the aforementioned separability principle.

The notion of separability has been viewed within the framework of classical physics as a principal condition of our conception of the world, a condition that characterizes all our thinking in acknowledging the physical identity of distant things, the "mutually independent existence (the 'being thus')" of spatiotemporally separated systems (Einstein 1948/1971, p. 169). The primary implicit assumption pertaining to this view is a presumed *absolute kinematic independence* between the knowing subject (the physical scientist) and the object of knowledge, or equivalently, between the measuring system (as an extension of the knowing subject) and the system under measurement. The idealization of the kinematically independent behavior of a physical system is possible in classical physics both due to the Cartesian-product structure of phase space, namely, the state-space of classical theories, and the absence of genuine indeterminism in the course of events or of an element of chance in the measurement process. During the act of measurement a classical system conserves its identity. Successive measurements of physical quantities, like position and momentum that define the state of a classical system, can be performed to any degree of accuracy and the results combined can completely determine the state of the system before and after the measurement interaction, since its effect, if not eliminable, takes place continuously in the system's state-space and is therefore predictable in principle.[1]

Consequently, classical physical quantities are taken to obey a so-called 'possessed values' or 'definite values' principle that may be succinctly formulated as follows:[2]

> *Definite values principle*: Any classical system is characterized, at each instant of time, by definite values for *all* physical quantities pertaining to the system in question.

That is, classical properties (values of physical quantities) are considered as being *intrinsic* to the system, as being *possessed* by the system itself. They are independent of whether or not any measurement is attempted on them and their definite values are independent of one another as far as measurement is concerned. Thus, the principle of value-definiteness implicitly incorporates the following assumption of non-contextuality:

> *Non-contextuality*: If a classical system possesses a property (value of a physical quantity), then it does so independently of any measurement context, i.e. independently of *how* that value is eventually measured.

This means that the properties possessed by a classical system depend in no way on the relations obtaining between it and a possible experimental or measurement context used to bring these properties about. If a classical system possesses a given property, it does so independently of possessing other values pertaining to other experimental arrangements. So, both 'value-definiteness' and 'non-contextuality' encapsulate the basic idea of the classical conception of physical reality, namely, its independence from its being measured.

*2.2. Classical Physical Ontology*

Hence, with respect to philosophical considerations, the aforementioned structural features of classical physics, most notably, separability and definite-values principle, give rise to a *classical physical ontology* that may be characterized by the following interrelated notions:

- *Classical realism*, asserting the absolute *metaphysical independence* of physical reality from the knowing subject, independence not in the trivial sense of considering physical reality as being there whether or not human observers exist, but in the sense of being *the way it is* whether or not it is observed and regardless of the acts or operations performed. This notion of realism consists in the assumption that whatever exists in the physical world is logically and conceptually independent of our measurements which serve to give us information about it. It is motivated by the classical idealization that the observed objects are indeed the entities in the world, the latter being capable of enjoying self-autonomous existence.
- *Classical objectivity*, assuming that objective knowledge of an object is achieved by forming a representation of that object as an entity possessing properties by itself. The objective meaning of these properties is tied to the fact that we conceive of them as corresponding to intrinsic properties of independently existing entities that enjoy intertemporal individuality in isolation from their environment. Specifically, the representational description of the object is considered as being independent of the means of describing it, including any experiments, measurements, etc. Thus, classical objectivity as an epistemological doctrine presupposes classical realism as an ontological doctrine.
- *Transcendent correspondence account of truth*, portraying a representational correspondence between the way *the world actually is* and the way *we observe the world to be*. In so far as classical concepts can be consistently viewed as referring to entities of the 'real world as it truly is', classical physical science, as a whole, may be, and indeed is systematically viewed by adherents of classical realism as providing an approximately accurate representation of the world as 'it really is'. The latter feature is reinforced by the fact that, according to the Boolean propositional structure of classical mechanics, the propositions of a classical system are semantically decidable (e.g., Dalla Chiara et al. 2004, p. 21); they are either determinately true or determinately false independently of our power to establish which value it is. Even if it is impossible to produce a basis on which we may ascertain the truth value of a proposition this does not imply that it does not possess any such value. It always has one. Consequently, the propositions of a classical system are regarded as possessing *determinate* truth values prior to and independent of any actual investigation of the states of affairs the propositions denote. That is, propositions in classical physics are considered as being either true or false in virtue of a stable and well-defined reality which serves as the implicit referent of every proposition, and thus as possessing an objective truth value regardless of our means of exploring and warranting its assignment.
- *Ontological reductionism*, according to which, a salient subset of the natural kind of entities inhabiting the world, together with their intrinsic properties and spatiotemporal relations they enter into, *fix* or *determine*, through a series of successive reductions, the nature and behavior of the universe as a whole. From this perspective, the world is viewed as being compartmentalized into atomic objects-systems, characterized by individual existence, and everything else supervenes upon them in conjunction with the spatiotemporal relations among them. If, therefore, one is able to determine the intrinsic properties of atomic objects in space and time, then one can describe the world completely. Evidently, this conception is closely related to the separability principle of classical physics. It is captured intuitively by the familiar fact that if one constructs a compound system or object by assembling its independently existing parts, then the physical properties of that system are completely determined by the properties of the parts and the way they have been combined so as to form the initial system of interest. This exemplification characterizes also, as a natural restriction to physics, David Lewis' ontological atomistic doctrine of Humean supervenience (e.g., Karakostas 2009a).[3]

## 3. The Quantum Conception of Nature

### 3.1. The Generalized Phenomenon of Non-Separability

In contrast to classical physics, standard quantum mechanics systematically violates the conception of separability.[4] From a formal point of view, the source of its defiance is due to the tensor-product structure of a compound Hilbert space and the quantum mechanical principle of the superposition of states, which incorporates a kind of *objective indefiniteness* for the numerical values of *any* observable belonging to a superposed state. The generic phenomenon of quantum non-separability, experimentally confirmed for the first time in the early 1980s, precludes in a novel way the possibility of defining individual objects independently of the conditions under which their behavior is manifested (e.g., Aspect et al. 1982; Tittel et al. 1998; Gröblacher et al. 2007). Even in the simplest possible case of a compound system $S$ consisting of just two subsystems $S_1$ and $S_2$ that have interacted at some time in the past, the compound system should be treated as a *non-separable*, *entangled* system, however large is the distance among $S_1$ and $S_2$. In such a case, it is not permissible to consider them individually as self-autonomous entities enjoying intertemporal identity. The global character of their behavior precludes any consistent description or any explanation in terms of individual systems, each with its own well-defined state or pre-determined physical properties. Only the compound system $S$, as a whole, is assigned a well-defined (non-separable) pure state. Therefore, when a compound system such as $S$ is in an entangled state, namely a superposition of pure states of tensor-product forms, *maximal determination of the whole system does not allow the possibility of acquiring maximal specification of its component parts*, a circumstance with no precedence in classical physics. In a paper related to the Einstein-Podolsky-Rosen argument, Schrödinger explicitly anticipated this counterintuitive state of affairs:

> When two systems, of which we know the states by their respective representations, enter into temporary physical interaction due to known forces between them, and then after a time of mutual influence the systems separate again, then *they can no longer be described in the same way as before, viz. by endowing each of them with a representative of its own*. ... I would not call that one but rather *the characteristic trait of quantum mechanics, the one that enforces its entire departure from classical lines of thought* (Schrödinger 1935/1983, p. 161; emphasis added).

The phenomenon of quantum non-separability undeniably reveals the holistic character of entangled quantum systems. Quantum mechanics is the first — and up to day the only — logically consistent, mathematically formulated and empirically well-confirmed theory, which incorporates as its basic feature that the 'whole' is, in a non-trivial way, more than the sum of its 'parts' including their spatiotemporal relations and physical interactions. Contrary to the situation in classical physics, when considering an entangled compound system, 'whole' and 'parts' are interconnected in such a way that their *bi-directional reduction* is, in principle, impossible (e.g., Karakostas 2009b). Intimately related to this, there exist properties considering any entangled quantum system which, in a clearly specifiable sense, characterize the whole system but are neither reducible to nor derived from any combination of local properties of its parts. As a means of exemplifying the preceding points, let us consider an important class of compound quantum systems that form the prototype of EPR-entangled systems, namely, spin-singlet pairs. Let then $S$ be a compound system consisting of a pair ($S_1$, $S_2$) of spin-1/2 particles in the following superposed state, known as the singlet state

$$W_S = 1/\sqrt{2}\ \{|\psi_+>_1 \otimes |\varphi_->_2 - |\psi_->_1 \otimes |\varphi_+>_2\}, \tag{1}$$

where $\{|\psi_\pm>_1\}$ and $\{|\varphi_\pm>_2\}$ are spin-orthonormal bases in the two-dimensional Hilbert spaces $H_1$ and $H_2$ associated with $S_1$ and $S_2$, respectively. As is well-known, in such a case, it is quantum mechanically predicted and experimentally confirmed that the spin components of $S_1$ and of $S_2$ have always opposite spin orientations; they are perfectly *anti-correlated*. Whenever the spin component along a given direction of, say, particle $S_1$ is measured at time $t_0$ and found equal to +1/2 $\hbar$ (correspondingly −1/2 $\hbar$), the subsequent destruction of the superposition bonds (between the tensor-product states involved) imparts to particle $S_2$ a tendency: that of inducing — in this special case, with certainty — the opposite value −1/2 $\hbar$ (correspondingly +1/2 $\hbar$), if and when, at a time $t>t_0$, particle $S_2$ is submitted to an appropriate measurement of the same component of spin as $S_1$. From a physical point of view, this derives from the interference (the anti-symmetric phase interrelations) with which the subsystem unit vectors $|\psi_\pm>_1$ and $|\varphi_\pm>_2$ — or, more precisely, the two product states $|\psi_+>_1 \otimes |\varphi_->_2$, $|\psi_->_1 \otimes |\varphi_+>_2$ — are combined within $W_S$. This, in turn, leads not only to the subsystem interdependence of the type described above but also to conservation of the total angular momentum for the pair ($S_1$, $S_2$) of spin-1/2 particles and thus to the property of definite total spin of value zero for the compound system $S$.

The latter is an *irreducible*, *holistic* property of $S$: it is not determined by any physical properties of its subsystems $S_1$, $S_2$ considered individually. Specifically, the property of $S$ 'having total spin zero' is not specified by the spin properties of $S_1$ and of $S_2$, since neither $S_1$ nor $S_2$ has any definite spin in the superposed singlet state $W_S$. Observe that in $W_S$, *no* pure spin state can be assigned to either of the particles $S_1$ and $S_2$, since neither of the corresponding unit vectors $|\psi_\pm>_1$ and $|\varphi_\pm>_2$ are eigenstates of $W_S$; the state $W_S$ is an eigenstate of the *total* spin operator $\sigma_1 \otimes I + I \otimes \sigma_2$, which cannot be understood as being composed of *definite individual* spin values of the two single particles. Hence, in the state $W_S$, *no* spin component of either particle $S_1$ or particle $S_2$ exists in an actual form, possessing *occurrent* spin properties. All three spin components of each particle, however, coexist in a potential form and any one component possesses the tendency of being actualized at the expense of the indiscriminacy of the others if the associated particle interacts with an appropriate measuring apparatus. In this respect, the spin-singlet state — as any compound superposed state — represents in essence the entanglement, the inseparable correlation of potentialities, whose content is not exhausted by a catalogue of actual pre-existing values that may be assigned to the spin properties of $S_1$ and $S_2$, separately.

Furthermore, the probability distributions concerning spin components of $S_1$ and of $S_2$ along some one direction do not ensure, with probability one, the property of $S$ 'having total spin zero'. Neither the latter property could be understood or accounted for by the possibility (that an adherent of reductionism may favor) of treating $S_1$ and $S_2$ separately at the expense of postulating a relation between them as to the effect of their spin components 'being perfectly anti-correlated'. For, whilst 'having total spin zero' is an intrinsic physical property of the compound system $S$ in the non-separable state $W_S$, the assumed relation is not an intrinsic physical relation that $S_1$ and $S_2$ may have in and of themselves. That is, although the relation of perfect anti-correlation is encoded within state $W_S$, ascribing this relation to individual parts of a system is not tantamount to being in state $W_S$. The relation of perfect anti-correlation is inherent to the entangled state $W_S$ itself which refers directly to the whole system. The entangled correlations between $S_1$ and $S_2$ just do not supervene upon any properties of the subsystem parts taken separately.

It may seem odd to consider non-supervenient relations holding between non-individuatable relata. However, the important point to be noticed is that within an entangled quantum system there is no individual pure state for a component subsystem alone. Within $W_S$ neither subsystem $S_1$ nor subsystem $S_2$ acquires individual independent existence. In considering any entangled compound system, the nature and properties of component parts may only be determined from their 'role' — the forming pattern of the inseparable web of relations — within the whole. Here, the part-whole relationship appears as complementary: the part is made 'manifest' through the whole, while the whole can only be 'inferred' via the interdependent behavior of its parts. Thus, in the example under consideration, the property of total spin of the whole in the singlet state $W_S$ does indicate the way in which the parts are related with respect to spin, although neither part possesses a definite numerical value of spin in any direction in distinction from the other one. And it is *only* the property of the total spin of the whole that contains *all* that can be said about the spin properties of the parts, because it is only the entangled state of the whole that contains the correlations among the spin probability distributions pertaining to the parts.[5] Consequently, the part-whole reduction with respect to the property of total spin zero in $W_S$ has failed: the latter property, whereas characterizes the whole system, is neither *reducible* to nor *supervenient* upon any properties of its constituent parts. Exactly the same holds for the properties of total momentum and relative distance of the overall system $S$ with respect to corresponding local properties of the parts. Analogous considerations, of course, to the aforementioned paradigmatic case of the spin-singlet pair of particles apply to any case of quantum entanglement. Entanglement need not be of maximal anti-correlation, as in the example of the singlet state.[6] It does neither have to be confined to states of quantum systems of the same kind; entanglement reaches in principle the states of all compound quantum systems.

This is precisely the delicate point with entangled correlations in Hilbert-space quantum mechanics: they cannot be reduced to or explained in terms of pre-assigned relations or interactions among the parts; their existence cannot be traced back to any interactions. Whereas the smallest interaction during the temporal development of the parts of a compound system gives rise to entanglement, entanglement itself needs no recourse to interaction for its being established. Interaction is a sufficient but not a necessary condition for entanglement. Quantum entanglement does occur in the absence of any interactions, since the origin of the phenomenon is essentially of a kinematical rather than dynamical nature, as dictated by the linear character of the superposition principle of states. Due to that the entangled correlations among the states of physical systems do not acquire the status of a causally determined relation.[7] Their delineation instead is specified by the entangled quantum state itself which refers directly to the whole system.

The generic phenomenon of quantum entanglement casts severe doubts on the existence of *isolated* (sub)systems and the applicability of the notion of atomism, in the sense that the parts of a quantum whole no longer exist as self-autonomous, intrinsically defined individual entities. The non-separable character of the behavior of an entangled quantum system precludes in a novel way the possibility of describing its component subsystems in terms of pure states. In fact, whenever the pure entangled state of a compound system is decomposed in order to represent subsystems, the effect can only extent up to a representation in terms of *incompletely specified* statistical (reduced) states of those subsystems (e.g., Blank et al. 1994). For, whenever a compound system is in an entangled state, as in Eq. (1), there are, in general, no pure states of the component subsystems on the basis of which the compound state of the whole system could be completely determined. Consequently, the legendary notion of the classical paradigm that the nature of the whole is fully describable or reducible to the properties of the parts is no longer defensible. In the framework of quantum mechanics, the state of the whole system cannot in general be determined by the states of its component parts, this being the case even when the parts occupy distinct regions of space however far apart. Because, at the quantum domain, it is exclusively only the compound state of the whole system that exhaustively specifies the probabilistic entangled correlations among the states of its parts. Hence, any case of quantum entanglement constitutes a violation of the separability principle, and the latter is the reason why entanglement induces a sort of holism in quantum mechanics.

*3.2. The Context-Dependence of Quantum Objects*

From a foundational viewpoint of quantum theory, the concept of quantum entanglement and the associated phenomenon of quantum non-separability refer to a context-independent, or in d' Espagnat's (2006) scheme, observer- or mind-independent reality. The latter is operationally inaccessible. It pertains to the domain of entangled correlations, potentialities and quantum superpositions obeying a non-Boolean logical structure. Here the notion of an object, whose aspects may result in intersubjective agreement, enjoys no a priori meaning independently of the phenomenon into which is embedded. In quantum mechanics in order to be able to speak meaningfully about an object, to obtain any kind of description, or refer to experimentally accessible facts the underlying wholeness of nature should be decomposed into interacting but disentangled subsystems. As will be argued in the sequel, in consonance with Primas (1993, 2007), well-defined separate objects (and their environments) are generated by means of a so-called Heisenberg cut (1958, p. 116), namely through the process of a deliberate abstraction/projection of the inseparable non-Boolean domain into a Boolean context that necessitates the suppression (or minimization) of entangled correlations between the object-to-be and the environment-to-be (e.g., a measuring apparatus).

The presuppositions of applying a Heisenberg cut are automatically satisfied in classical physics, in conformity with the separability principle of Section 2.1. In a non-separable theory like quantum mechanics, however, the concept of the Heisenberg cut acquires the status of a methodological regulative principle through which access to empirical reality is rendered possible. The innovation of the Heisenberg cut, and the associated separation of a quantum object from its environment, is mandatory for the description of measurements. It is, in fact, necessary for the operational account of any directly observable pattern of empirical reality. The very possibility of devising and repeating a controllable experimental procedure presupposes the existence of such a subject-object separation. Without it the concrete world of material facts and data would be ineligible; it would be conceived in a totally entangled manner. In this sense, a physical system may account as an experimental or a measuring device only if it is not holistically correlated or entangled with the object under measurement.

Consequently, any atomic fact or event that 'happens' is raised at the empirical level only in conjunction with the specification of an experimental arrangement[8] — an experimental context that conforms to a Boolean domain of discourse — namely to a set of observables co-measurable by that context. In other words, there cannot be well-defined events in quantum mechanics unless a specific set of co-measurable observables has been singled out for the system-experimental context whole (e.g., Landsman 1995). For, in the quantum domain, one cannot assume, without falling into contradictions, that observed objects enjoy a separate well-defined identity irrespective of any particular context. One cannot assign, in a consistent manner, definite sharp values to all quantum mechanical observables pertaining to a microscopic object, in particular to pairs of incompatible observables, independently of the measurement context actually specified. In terms of the structural component of quantum theory, this is due to functional relationship constraints that govern the algebra of quantum mechanical observables, as revealed by the Kochen-Specker (1967) theorem and its recent investigations (e.g., Mermin 1995, Cabello et al. 1996, Hasegawa et al. 2006, Kirchmair et al. 2009). In view of them, it is not possible, not even *in principle*, to assign to a quantum system non-contextual properties corresponding to *all* possible measurements. This means that it is not possible to assign a definite unique answer to every single yes-no question, represented by a projection operator, independent of which subset of mutually commuting projection operators one may consider it to be a member. Hence, by means of a generalized example, if *A*, *B* and *C* denote observables of the same quantum system, so that the corresponding projection operator *A* commutes with operators *B* and *C* ($[A, B] = 0 = [A, C]$), not however the operators *B* and *C* with each other ($[B, C] \neq 0$), then the result of a measurement of *A* *depends* on whether the system had previously been subjected to a measurement of the observable *B* or a measurement of the observable *C* or in none of them. Thus, the value of the observable *A* depends upon the set of mutually commuting observables one may consider it with, that is, the value of *A* depends upon the selected set of measurements. In other words, the value of the observable *A* cannot be thought of as pre-fixed, as being independent of the experimental context actually chosen, as specified, in our example, by the *B* or *C* frame of mutually compatible observables. In fact, any attempt of simultaneously attributing context-independent, sharp values to all observables of a quantum object forces the quantum statistical distribution of value assignment into the pattern of a classical distribution, thus leading directly to contradictions of the Greenberger-Horne-Zeilinger type (for a recent discussion see Greenberger 2009).

This state of affairs reflects most clearly the unreliability of the so-called 'definite values' or 'possessed values' principle of classical physics of Section 2.1, according to which, values of physical quantities are regarded as being possessed by an object independently of any measurement context. The classical-realist underpinning of such an assumption is conclusively shown to be incompatible with the structure of the algebra of quantum mechanical observables. Well-defined values of quantum observables can, in general, be regarded as pertaining to an object of our interest only within a framework involving the experimental conditions. The latter provide the necessary conditions whereby we make meaningful statements that the properties attributed to quantum objects are part of physical reality. Consequent upon that the exemplification of quantum objects is a *context-dependent* issue with the experimental procedure supplying the physical context for their realization. The introduction of the latter operates as a formative factor on the basis of which a quantum object manifests itself. The classical idealization of sharply individuated objects possessing intrinsic properties and having an independent reality of their own, a self-autonomous existence, breaks down in the quantum domain. Quantum mechanics describes physical reality in a substantially context-dependent manner.

Accordingly, well-defined quantum objects cannot be conceived of as 'things-in-themselves', as 'absolute' bare particulars of reality, enjoying *intrinsic* individuality or *intertemporal* identity. Instead, they represent carriers of patterns or properties which arise in interaction with their experimental context/environment, or more generally, with the rest of the world;[9] the nature of their existence — in terms of state-property ascription — depends on the context into which they are embedded and on the subsequent abstraction of their entangled correlations with the chosen context of investigation. Thus, the resulting contextual object *is* the quantum object exhibiting a particular property with respect to a certain experimental situation. The contextual character of property-ascription implies, however, that a state-dependent property of a quantum object is not a well-defined property that has been possessed *prior* to the object's entry into an appropriate context. This also means that not all contextual properties can be ascribed to an object at once. One and the *same* quantum object does exhibit several possible contextual manifestations in the sense that it can be assigned several definite incommensurable properties only with respect to distinct experimental arrangements which mutually exclude each other. Thus, in contradistinction to a mechanistic or naive realistic perception, we arrive at the following general conception of an object in quantum mechanics. According to this, a quantum object — as far as its state-dependent properties are concerned — constitutes a totality defined by all the possible relations in which this object may be involved. Quantum objects, therefore, are viewed as carriers of inherent dispositional properties. In conjunction with our previous considerations of Section 3.1, ascribing a property to a quantum object means recognizing this object an *ontic potentiality* to produce effects whenever it is involved in various possible relations to other things or whenever it is embedded within an appropriate experimental context.

Consequently, a quantum object is not an individual entity that possesses well-defined intrinsic properties at all times even beyond measurement interactions, nor is it a well-localized entity in space and time that preserves deterministic causal connections with its previous and subsequent states, allowing it, thereby, to traverse determinate trajectories.[10] In fact, a quantum object exists, independently of any operational procedures, only in the sense of 'potentiality', namely, as being characterized by a set of potentially possible values for its various physical quantities that are actualized when the object is interacting with its environment or a pertinent experimental context.[11] Due to the genuinely non-separable structure of quantum mechanics and the subsequent context-dependent description of physical reality, a quantum object can produce *no* informational content that may be subjected to experimental testing without the object itself being transformed into a contextual object. Thus, whereas quantum non-separability refers to an *inner-level* of reality, a mind-independent reality that is operationally elusive, the introduction of a context is related to the *outer-level* of reality, the contextual or empirical reality that results as an abstraction in the human perception through deliberate negligence of the all-pervasive entangled (non-separable) correlations between objects and their environments. In this sense, quantum mechanics has displaced the verificationist referent of physics from 'mind-independent reality' to 'contextual' or 'empirical reality'.

### 3.3. Levels of a Unified Reality: From Mind-Independent to Empirical Reality

Strictly speaking the concept of a mind-independent reality is not purely scientific; it does not constitute a matter of physics or mathematics; it is rather metaphysical by nature. It concerns, by definition, the existence of things in themselves regardless of any act of empirical testing. Consequently, it does not apply to empirical science proper. It may be viewed, however, as a regulative principle in physics research, as a conviction which gives direction and motive to the scientific quest. As Einstein put it:

> It is basic for physics that one assumes a real world existing independently from any act of perception. But this we do not *know*. We take it only as a programme in our scientific endeavors. This programme is, of course, prescientific and our ordinary language is already based on it (quoted in Fine 1996, p. 95).

Granting the metaphysical or heuristic character of its nature, we nonetheless consider the notion of a mind-independent reality as unassailable in any scientific discourse; we amply recognize its existence as being logically *prior* to experience and knowledge; we acknowledge its external to the mind structure as being responsible for *resisting* human attempts in organizing and conceptually representing experience.

But, significantly, in the quantum domain, the nature of this independent reality is left unspecified. For, due to the generalized phenomenon of quantum non-separability, we must conceive of independent reality as a highly entangled whole with the consequence that it is impossible to conceive of parts of this whole as individual entities, enjoying autonomous existence, each with its own well-defined pure state. Neither reality considered as a whole could be comprehended as the sum of its parts, since the whole, according to considerations of Section 3.1, cannot be reduced to its constituent parts in conjunction with the spatiotemporal relations among the parts. Quantum non-separability seems to pose, therefore, a novel limit on the ability of scientific cognizance in revealing the actual character of independent reality itself, in the sense that any detailed description of the latter necessarily results in *irretrievable* loss of information by dissecting the otherwise undissectable. From a fundamental viewpoint of quantum mechanics, any discussion concerning the nature of this indivisible whole is necessarily of an ontological, metaphysical kind, the only confirmatory element about it being the network of entangled interrelations which connect its events. In this respect, it can safely be asserted that reality thought of as a whole is not scientifically completely knowable, or, at best, in d' Espagnat's (2006) expression, it is veiled. Hence, our knowledge claims to reality can only be partial, not total or complete, extending up to the structural features of reality that are approachable by the penetrating power of the theory itself and its future development.[12]

The term 'reality' in the quantum realm cannot be considered to be determined by what physical objects really are in themselves. As already argued, this state of affairs is intimately associated with the fact that, in contrast to classical physics, values of quantum mechanical quantities cannot, in general, be attributed to a quantum object as intrinsic properties. Whereas in classical physics, nothing prevented one from considering *as if* the phenomena reflected intrinsic properties, in quantum physics, even the *as if* is precluded. Indeed, quantum phenomena are not stable enough across series of measurements of non-commuting observables in order to be treated as *direct reflections* of invariable properties. The values assigned to quantum mechanical observables cannot be said to belong to the observed object alone regardless of the overall experimental context which is relevant in any particular situation. Hence, well-defined quantum objects, instead of picturing entities populating the mind-independent reality, they depict the *possible manifestations* of these entities within a concrete experimental context. In this respect, the quantum mechanical framework seems only to allow a detailed description of reality that is co-determined by the projection of reality into a particular context. Without prior information of the kind of observables used to specify a context and thus to prepare a quantum mechanical state, it is just not possible to find out what the *actual* state of a quantum system is; measurement of observables that do not commute with this original set will inevitably produce a different state. What contemporary physics, especially quantum mechanics, can be expected therefore to describe is not 'how mind-independent reality is', as classical physics may permit one to presume. Within the domain of quantum mechanics, knowledge of 'reality in itself', 'the real such as it truly is' independent of the way it is contextualized, is impossible in principle.[13] Thus, it is no longer conceivable to judge the reliability of our knowledge through a comparison with reality itself, and in the scientific description we must adopt alternative necessary conditions for meeting a suitable criterion of objectivity.

*3.4. Object and Objectivity*

To this end we underline the fact that although contextual objects cannot be viewed, by definition, as objects in an absolute, intrinsic sense, nonetheless, they preserve *scientific objectivity*; they reflect structures of reality in a manner that is independent of various observers or of any observer's cognition. For, since they are given at the expense of quantum mechanical non-separability, the 'conditions of their being experienced' are determined by the 'conditions of accessibility', or more preferably, in reinterpreting Cassirer (1936/1956, p. 179) in the above expression, by the '*conditions of disentanglement*'. Once the latter conditions are specified, the result of their reference is intersubjective since it is valid for any observer whatsoever. In other words, given a particular experimental context, concrete objects (structures of reality) have well-defined properties independently of our knowledge of them. Thus, within the framework of quantum mechanics, the perceptible separability and determinateness of the contextual objects of empirical reality are generated by means of an experimental intervention that suppresses (or sufficiently minimizes) the factually existing entangled correlations of the object concerned with its environment. It is then justified to say that the fulfillment of disentanglement conditions provides a level of description to which one can associate a separable, albeit contextual, concept of reality whose elements are objectively experienced as distinct, well-localized objects having determinate properties.

Furthermore, since the contextual object constitutes the actually determinable appearance of the quantum object, quantum objects are *objectively real* in the sense that they are manifested to us in the context of lawful connections; they also contribute to the creation of such lawful connections. Hence, we are confronted in the quantum domain with a reversal of the classical relationship between the concepts of object and law, a situation that has been more vividly expressed in broader terms (of a neo-kantian type, not necessarily adopted here in toto) by Cassirer. In his words:

> ... objectivity itself — following the critical analysis and interpretation of this concept — is only another label for the validity of certain connective relations that have to be ascertained separately and examined in terms of their structure. The tasks of the criticism of knowledge ("Erkenntniskritik") is to work backwards from the unity of the general object concept to the manifold of the *necessary and sufficient conditions that constitute it*. In this sense, that which knowledge calls its "object" breaks down into *a web of relations* that are held together in themselves through the highest rules and principles (Cassirer 1913, transl. in Ihmig 1999, p. 522; emphasis added).

Although Cassirer's reference is within the context of relativity theory, where these 'highest rules and principles' stand for the symmetry principles and transformations which leave the relevant physical quantities invariant, in the quantum domain, a *pre-condition* of something to be viewed as an object of scientific experience is the elimination of the entangled correlations with its environment. In other words, in order for any object-system *S* of the quantum realm, its observed qualities (e.g., any obtainable measuring results on *S*) to be considered as properties of *S*, the condition of disentanglement must be fulfilled. Thus, disentanglement furnishes a *necessary condition* for a quantum object to become amenable to scientific analysis and experimental investigation; that is, disentanglement constitutes a necessary material pre-condition of quantum physical experience by rendering the object system *S* a scientific object of experience. This marks an essential difference between merely believing in the existence of objects, and being aware of the procedure through which scientific objects of experience are constituted.

In this respect, disentanglement may be viewed as a *background constitutive principle* — as a *pre-condition* of quantum physical experience — which is *necessary* if quantum mechanics is to grasp empirical reality at all. This kind of (conditional) necessity, however, cannot be taken in an absolute sense as, for instance, in Kant's timeless a priori necessary forms of the possibility of all experience. But rather in the restrictive sense that given at a historical period a theoretical project of well-confirmed knowledge, namely, in our case, quantum mechanics, disentanglement forms a pre-condition of any possible access to (empirical) reality, of any possible empirical inquiry on the microscopic scale, and hence of any possible cognizance of microphysical objects as scientific objects of experience.

## 4. Viewing the World from Within

In light of the preceding considerations, the common philosophical assumption concerning the feasibility of a panoptical, Archimedean point of view is rendered illusory in quantum mechanics. In contrast to an immutable and universal 'view from nowhere' of the classical paradigm, quantum mechanics acknowledges in an essential way a perspectival/contextual character of knowledge. Although possible in classical physics, in quantum mechanics we can no longer display the whole of nature in one view. As argued in Section 3.2, access to the non-Boolean quantum world is only gained by adopting a particular Boolean perspective, by specifying a certain Boolean context which breaks the wholeness (the underlying non-separability) of nature. Consequently, the description and communication of results of experiments in relation to the non-Boolean quantum world presuppose the impossibility of a perspective-independent account, since one must at the outset single out an experimental context (determined by a set of co-measurable observables for the context-cum-quantum system whole) and in terms of which the definite result of a measurement can be realized.

Be sure there is only one external reality, but every description of it presupposes — according to quantum mechanics — the adoption of a particular point of view. There is no such a thing as a 'from nowhere' perspective, envisaged by Nagel (1986), or a universal viewpoint. We are part of the world we explore and therefore we cannot view it from 'without'; our observations are always conditioned upon the fact that we are 'in' the world; we cannot transcend this limitation. A complete knowledge of the world as a whole would have to provide an explanation of the conditions for description and communication which we ourselves, as cognizant subjects, are already subjected to. It would have to include within a hypothetically posited ultimate theory an explanation of the knowing subject and his pattern recognition mechanisms. This would be like attempting to produce a map of the globe which contained itself as an element. The usage of this metaphor is meant to convey the conceptually deep fact that a logically consistent theory cannot generally describe its universe as its own object. In particular, the scientific language of our hypothetical universal ultimate theory would have to be semantically closed, and hence engender antinomies or paradoxes especially in relation to self-referential descriptions, as in the case of von Neumann's account of quantum measurement that leads to an infinite regress of observing observers.[14]

Be that as it may, the assumption of a 'view from nowhere' appeared realizable prior to quantum mechanics, because in classical physics the validity of separability and unrestricted causality led to the purely reductionist presumption that one could consistently analyze a compound system into parts and consequently deduce the nature of the whole from its parts. Since the part could be treated as a closed system separate from the whole, the whole could ultimately be described — by applying the conservation laws of energy, momentum and angular momentum — as the sum of its parts and their physical interactions, and hence the knowing subject would achieve its knowledge of physical reality from the 'outside' of physical systems.

In the quantum theoretical framework that picture is no longer valid. As we have extensively argued, the consideration of physical reality as a whole — supposedly that a sense is ascribed to this word — cannot be conceived of as the sum of its parts in conjunction with the spatiotemporal relations or physical interactions among the parts, since the quantum whole provides the framework for the existence of the parts. In considering any case of quantum entanglement, the interrelation between the parts cannot possibly be disclosed in an analysis of the parts that takes no account of the entangled connection of the whole. As already shown, their entangled relation does not supervene upon any intrinsic or relational properties of the parts taken separately. This is indeed the feature which makes the quantum theory go beyond any mechanistic or atomistic thinking. In consistency with classical physics' atomistic picture, given any compound physical system, the intrinsic properties of the whole were regarded as being reducible to or supervenient upon the properties of its parts and their spatiotemporal relations. In quantum mechanics the situation is actually reversed; due to the genuinely non-separable structure of quantum theory, the state-dependent properties of the parts can ultimately be accounted only in terms of the characteristics of the whole. In a truly non-separable physical system, as in an entangled quantum system, the part does acquire a different identification within the whole from what it does outside the whole, in its own 'isolated', separate state (see esp. Section 3.1). Thus, for instance, no definite spin property of an isolated spin-1/2 particle, e.g. a 'free' or 'bare' electron, can be identified with the spin property of either member of a pair of electrons in the singlet state, since in this situation any spin state can be specified only at the level of the overall system. When in the singlet state, there is simply no individual (pure) spin state for a component particle alone, unless explicit reference is made to the partner particle via the total information contained in the compound state. Consequently, any spin state of either particle is fixed only through the interconnected web of entangled relations among the particles. Hence, the spin property of either particle, when in an entangled state, cannot stand alone, enjoying self-autonomous existence, independently of the interrelations within the whole.

When all is said and done, the present situation in physics suggests that the natural scientist as a conscious being may operate within a mild form of the reductionist paradigm in trying to analyze complex objects in terms of parts with the absolute certainty, however, that during the process the nature of the whole will not be disclosed. The value of the reductionistic concept as a working hypothesis or as a methodological tool of analysis and research is not jeopardized at this point,[15] but *ontologically* it can no longer be regarded as a true code of the actual character of the physical world and its contents. Quantum mechanical non-separability strongly suggests that the functioning of the physical world cannot just be reduced to that of its constituents thought of as a collection of interacting but separately existing localized objects. Any coherent conceptualization of the physical world that is compatible with the predictions of quantum mechanics requires us to view the world, in the expression of Heisenberg (1958, p. 96), "as a complicated tissue of events, in which connections of different kinds alternate or overlay or combine and thereby determine the texture of the whole". Although the latter can hardly be fully knowable, an enlightenment of its actual character may be given by the penetrating power of the theory itself and its future development. In this respect, it is rather safe to conjecture that the conception of quantum non-separability will be an integral part of the next conceptual revolution in physics and may even be used as a regulative constructive hypothesis guiding the search for our deeper understanding of nature.

## 5. Concluding Remarks

In closing this work, and in view of its context, I wish to promote a spirit of tolerance with regard to realism/antirealism debate. On the one hand, let the advocates of strict realism refrain from claiming as absolute dogma that what is real must be comprehensible in its totality; since this is not supported — in fact, as already shown, it is rejected — by science itself. A projection of science, especially of physics, as being able to reveal — even in principle — the ultimate truth is a false image of science; it is deficient to really understand reality. A consistent understanding of modern science and its practice requires that we give up the idea that science aims at the description of reality as it truly is in itself. Neither contemporary physical science lends support to an a priori subject-object partition, an absolute dichotomy between the knowing subject and the object to be known, allegedly furnishing a perspective-free account of the world. Quantum mechanics reveals that the hunt of a universal perspective for describing physical reality is in vain. In the quantum domain of inquiry, it would be illusory to search for an overall frame by virtue of which one may utter 'this' or 'that', 'really is' independently of a particular context of reference. It is probably one of the deepest insights of contemporary quantum theory that whereas the totality of all experimental facts can only be represented on the basis of a globally non-Boolean theory, the acquisition of every single fact depends on a locally Boolean context. Furthermore, no two or more mutually incompatible contexts could be co-joined so as to provide a full picture of reality at any temporal moment. In quantum mechanical considerations summing up perspectives does not eliminate the inclusion of perspectives, does not result in no perspective at all. Accordingly, a non-contextual realist interpretation of modern physics is inappropriate. Quantum mechanical reality is non-separable and her distinctiveness into facts a matter of context. Yet, quantum non-separability does not contradict an objective, realist view of the world; it rather points to the abandonment of the classical conception of physical reality and its traditional philosophical presuppositions. After all, quantum mechanical features like value-indefiniteness, superposition, entanglement/non-separability, contextuality are connected in an inextricable way within quantum theory, so that their absence from the worldview of classical physics is no coincidence.

On the other hand, let the followers of the antirealist camp avoid condemning any inclination to deal with reality as an idle metaphysical exercise. Physics is not confined to purely operational, descriptive accounts of things; the consideration of existential questions does not entirely fall outside its realm, since the real difficulty lies in the fact that, in the words of Einstein,[16] "physics is a kind of metaphysics; physics describes 'reality'. But we do not know what 'reality' is; we know it only by means of the physical description!".

**Notes**

[1] It should be noted that this is hardly the case in the quantum theory of the measurement process (e.g., Karakostas & Dickson 1995).

[2] The principle of value-definiteness has variously been called in the literature as, for instance, "the determined value assumption" in Auletta (2001, pp. 21, 105).

[3] Undoubtedly, the classical physical ontology presented here is fully compatible with the structure of classical physics. Nonetheless, it is still an interpretation of classical physics, rather than being uniquely dictated by it, as after all the historical record of Kant's and Mach's reinterpretation of mechanics shows.

[4] In this work we shall not consider in any detail alternative interpretations to Hilbert-space quantum mechanics as, for instance, Bohm's ontological or causal interpretation.

[5] In this connection see Esfeld (2004). Also Rovelli (1996) and Mermin (1998) highlight the significance of correlations as compared to that of correlata.

[6] It is well known that spin-singlet correlations violate Bell's inequalities. We note in this connection the interesting result of Gisin (1991), Popescu & Rohrlich (1992) that for *any* entangled state of a two-component system there is a proper choice of pairs of observables whose correlations do violate Bell's inequality.

[7] For instance, the entangled correlations between spatially separated systems cannot be explained by assuming a direct causal influence between the correlated events or even by presupposing the existence of a probabilistic common cause among them in Reichenbach' s sense. Butterfield (1989) and van Fraassen (1989) have shown that such assumptions lead to Bell's inequality, whereas, as well-known, the latter is violated by quantum mechanics. See in addition, however, Szabó & Redei (2004) for the notion of a *common* common cause in relation to quantum correlations among spatially separated events.

[8] It should be pointed out that Bohr already on the basis of his complementarity principle introduced the concept of a 'quantum phenomenon' to refer "exclusively to observations obtained under specified circumstances, including an account of the whole experiment" (Bohr 1963, p. 73). This feature of context-dependence is also present in Bohm's ontological interpretation of quantum theory by clearly putting forward that "quantum properties cannot be said to belong to the observed system alone and, more generally, that such properties have no meaning apart from the total context which is relevant in any particular situation. In this sense, this includes the overall experimental arrangement so that we can say that measurement is context dependent" (Bohm & Hiley 1993, p. 108).

[9] Note that the so-called invariant or state-independent, and therefore, context-independent properties — like 'rest-mass', 'charge' and 'spin' — of elementary objects-systems can only characterize a certain class of objects; they can only specify a certain sort of particles, e.g., electrons, protons, neutrons, etc. They are not sufficient, however, for determining a member of the class as an individual object, distinct from other members within the same class, that is, from other objects having the same state-independent properties. Thus, an 'electron', for instance, could not be of the particle-kind of 'electrons' without fixed, state-independent properties of 'mass' and 'charge', but these in no way suffice for distinguishing it from other similar particles or for 'individuating' it in any particular physical situation. For a detailed treatment of this point, see, for example, French & Krause (2006).

[10] In standard quantum mechanics, it is not possible to establish a causal connection between a property $A(t)$ at time t and the same property $A(t'')$ at a later time $t''$, both pertaining to an object-system $S$, if $S$ had been subjected at a time value $t'$, $t<t'<t''$, to a measurement of a property $B$ incompatible with $A$. Because the successive measurement of any incompatible property of this kind would provide an uncontrollable material change of the state of $S$. Thus, a complete causal determination of all possible properties of a quantum object, most notably, coordinates of position and their conjugate momenta, allowing the object, henceforth, to traverse well-defined trajectories in space-time is not possible.

[11] The view that the quantum state vector refers to 'possibilities' or 'tendencies', as a certain extension of the Aristotelian concept of 'potentia', has been advocated by Heisenberg (1958, pp. 42, 53) in his later writings on the interpretation of quantum mechanics, and especially by Fock (1957, p. 646). Margenau (1950, pp. 335-337, 452-454) too has used the concept of 'latency' to characterize the indefinite quantities of a quantum mechanical state that take on specified values when an act of measurement forces them out of indetermination. Analogous is Popper's (1990, ch.1) understanding of attributing properties to quantum systems in terms of objective 'propensities'. Today one of the most eloquent defenders of the appropriateness of the concept of potentiality in interpreting quantum mechanics is Shimony (1993, Vol. 2, ch. 11), whereas a systematic development of the dialectical scheme 'potentiality-contextuality' for interpreting quantum mechanics is given in Karakostas (2007).

[12] This claim should not be conflated with the thesis of ontic structural realism set out notably by Steven French and James Ladyman (e.g., French and Ladyman 2003), and, according to which, only

structures in the sense of relations that are instantiated in the world are real; on this view, objects standing in the relations are simply non-existent (ibid. pp. 41-42). Our main objection against the thesis of ontic structural realism is that it dispenses altogether with physical objects. For, concrete relations that are instantiated in the natural world presuppose relata, that is, objects among which the relations obtain and of which they are predicated. What is challenging about quantum physics is not that there are no objects, but that the properties of quantum objects are remarkably different from the properties that classical physics considers. For instance, in any case of quantum entanglement, conceived as a relation among quantum objects, there are no intrinsic properties of the objects concerned on which the relation of entanglement obtains (see Section 3.1). The fact, however, that quantum objects cannot be individuated, in the classical sense, does not imply their inexistence. In other words, the non-individuality of quantum objects is not and cannot be tantamount to pronouncing their non-existence.

[13] It is tempting to think that a similar sort of context-dependence already arises in relativity theory. For instance, if we attempt to make context-independent attributions of simultaneity to spatially distant events — where the context is now determined by the observer's frame of reference — then we will come into conflict with the experimental record. However, given the relativization of simultaneity — or the relativization of properties like length, time duration, mass, etc. — to a reference frame of motion, there is nothing in relativity theory that precludes a complete description of the way nature is. Within the domain of relativity theory, the whole of physical reality can be described from the viewpoint of any reference frame, whereas, in quantum mechanics such a description is inherently incomplete.

[14] On this perspective, the insurmountable difficulties encountered in a complete description of the measuring process in quantum mechanics may not be just a flaw of quantum theory, but they may arise as a logical necessity in any theory which contains self-referential aspects, as it attempts to describe its own means of verification. Whereas the measuring process in quantum mechanics serves to provide operational definitions of the mathematical symbols of the theory, at the same time, the measurement concept features in the axiomatic structure of the theory, and the requirement of it's being described in terms of the theory itself induces a logical situation of semantical completeness which is reminiscent of Gödel's (1931/1962) undecidability theorem; the consistency of a system of axioms cannot be verified because there are mathematical statements that can neither be proved nor disproved by the formal rules of the theory; nonetheless, they may be verified by meta-theoretical reasoning.

[15] The sense of holism appearing in quantum mechanics, as a consequence of non-separability, should not be regarded as the opposite contrary of methodological reductionism. Holism is not an injunction to block distinctions. A successful holistic research program has to account apart from non-separability, wholeness and unity also for part-whole differentiation, particularity and diversity. In this respect, holism and methodological reductionism appear as complementary viewpoints, none can replace the other, both are necessary, none of them is sufficient.

[16] Einstein's letter to Schrödinger, 19 June 1935, cited in Howard (1989, p. 224).